\documentclass[aps,preprint]{revtex4}
\usepackage{amssymb,epsf}
\usepackage{amsmath}
\usepackage{graphicx}

\begin{document}

\title{Thermodynamics of Lovelock-Lifshitz Black Branes}
\author{M. H. Dehghani$^{1,2}$\footnote{mhd@shirazu.ac.ir} and R. B. Mann$^1$\footnote{rbmann@sciborg.uwaterloo.ca}}
\affiliation{$^1$ Department of Physics and Astronomy, University of Waterloo, 200 University
Avenue West, Waterloo, Ontario, Canada, N2L 3G1\\
$^2$ Physics Department and Biruni Observatory,
College of Sciences, Shiraz University, Shiraz 71454, Iran}
\begin{abstract}
We investigate the thermodynamics of Lovelock-Lifshitz black branes.
We begin by introducing the finite action of third order Lovelock gravity in the
presence of a massive vector field for a flat boundary, and use it to compute the energy density of these black branes. Using the field equations, we find a conserved quantity along the $r$ coordinate that relates the metric parameters at the horizon and at infinity. Remarkably, though
the subleading large-$r$ behavior of Lovelock-Lifshitz black branes differs substantively from their Einsteinian Lifshitz counterparts,  we find that the relationship between the energy density, temperature, and entropy density is unchanged from Einsteinian gravity. Using  the first law of thermodynamics to obtain the relationship between entropy and temperature, we find that it too is
the same as the Einsteinian case, apart from a constant of integration that depends on the Lovelock coefficients.
\end{abstract}
\maketitle

\section{Introduction}
Gauge/gravity duality presents a powerful new framework with which one
may study strongly coupled gauge theories. Indeed, it has been conjectured that
there is a duality between certain strongly coupled gauge theories and weakly coupled
string theories, which means that both theories  describe the same physics,
but calculations may be easier in one theory than the other.
More specifically, the AdS/CFT correspondence relates ${\mathcal N}=4$ supersymmetric
SU(N) gauge theory to superstrings in 10 dimensions \cite{Mal,Wit}.
One typical and fascinating aspect of  gauge/gravity duality is the
property of holography, which states that the
amount of information contained in the boundary gauge theory
is the same as the one contained in the bulk string theory.
This holographic toolbox has since spread its usage to many branches
of physics from hydrodynamic \cite{Kov} to condensed matter systems \cite{Hart}.
However these studies have been concerned with relativistic theories with a conformal symmetry
in the ultraviolet, which are described by asymptotically Anti-de Sitter (AdS) spacetimes.
Applying gauge/gravity duality to condensed matter theories with anisotropic scaling symmetry,
\begin{equation}
t\rightarrow \lambda ^{z}t,\hspace{1cm}\mathbf{{x}\rightarrow \lambda {x}}
\label{Ssym}
\end{equation}
where z is the dynamical exponent, suggests that we can further extend holography to encompass
nonrelativistic and Lifshitz field theories.

From a holographic point of view, this suggests the following (asymptotic)
form for the spacetime metric
\begin{equation}
ds^{2}=L^{2}\left( -r^{2z}dt^{2}+\frac{dr^{2}}{r^{2}}+r^{2}d\mathbf{x}%
^{2}\right)  \label{Lifmet}
\end{equation}
that obeys the scale invariance
\begin{equation}
t\rightarrow \lambda ^{z}t,\hspace{0.5cm}r\rightarrow \lambda ^{-1}r,\hspace{%
0.5cm}\mathbf{{x}\rightarrow \lambda {x}.}  \label{Ssym2}
\end{equation}
noted previously in a braneworld context \cite{Koroteev}.
A four-dimensional anisotropic scale invariant background using an action
involving a two form and a three form field with a Chern-Simons coupling
or a massive vector field can be engineered to yield solutions with this asymptotic behavior \cite
{Kach}. For these matter fields, lots of
effort has been expended in extending this solution to the case of
asymptotic Lifshitz solutions. One of the first analytic examples was
reported in Ref. \cite{Tay} for a sort of higher-dimensional dilaton gravity
without restricting the value of the dynamical exponent $z$. An exact
topological black hole solution that is
asymptotically Lifshitz with $z = 2$ was obtained in \cite{Mann}; further
solutions with $z = 4$ and with spherical topology were subsequently
obtained \cite{Peet1}. In general, however, such asymptotic Lifshitz black
holes must be obtained numerically \cite{Mann,Peet1,Bal1}. Asymptotic Lifshitz solutions in the vacuum of
higher-derivative gravity theories (with curvature-squared terms in the
action) have been investigated \cite{Ay}; the higher-curvature terms with
suitable coupling constant play the role of the desired matter. Recently, we introduced
some solutions -- both analytically and
numerically -- which can be regarded as higher-curvature modifications from Lovelock gravity
to those obtained from Einsteinian gravity coupled to matter \cite{DM}.

The holography of gravity theories including higher powers of the curvature, particularly
curvature-cubed interactions, have attracted increased attention \cite{Boer,Myers}.
This is because, in the context of AdS/CFT correspondence, corrections from higher powers of the
curvature must be considered on the gravity side of the correspondence in order to investigate CFTs with different values of their central charges.
On the other side, the asymptotic Lifshitz black hole from a holographic point of view needs
an extension of holography to encompass nonrelativistic field theories.  Our aim here is
to further develop the holographic dictionary for asymptotically
Lifshitz spacetimes in Lovelock gravity by introducing the counterterm method for   Lovelock-Lifshitz black branes \cite{DM} with flat boundary. Indeed, the generalization of holographic
techniques to this new context may offer new insights into the nature of the relation
between quantum gravity in asymptotically non-AdS spacetimes and the dual field
theory.  Here we generalize the counterterm method introduced
in \cite{Ross} for four-dimensional Einstein gravity to ($n+1$)-dimensional Lovelock
gravity, which of course contains the higher-dimensional Einstein gravity and
Gauss-Bonnet gravity too. The counterterm method for asymptotic AdS solutions of Lovelock gravity with flat boundary was introduced by us in \cite{DM2}. We generalize it to include asymptotic Lifshitz solutions of Lovelock gravity.

We make use of these generalizations to investigate the thermodynamics of Lovelock-Lifshitz black branes \cite{DM}.  Somewhat remarkably we find that the relationship between the energy density, temperature, and entropy density is unchanged from Einsteinian gravity.  We make use of the first law of thermodynamics to obtain the relationship between entropy and temperature.  We find this is also the same as the Einsteinian case, apart from a constant of integration that depends on the parameters in the Lovelock action.

The outline of our paper is as follows. We introduce the one dimensional Lagrangian
of the Lovelock-Lifshitz black branes in Sec. \ref{Lag}. We also write down the field equations and present
the constant $\mathcal{C}_{0}$ which is preserved along the coordinate $r$. This constant relates the information
at the horizon and at infinity. In Sec. \ref{finite}, we generalize the counterterm method introduced
 for the four-dimensional Einstein equation \cite{Ross} to Lovelock gravity, and present the
finite action for Lovelock-Lifshitz black branes. Section \ref{Constant} is devoted to the calculation of
the constant $\mathcal{C}_{0}$ at the horizon and at infinity. In Sec. \ref{Therm},
we obtain the energy density of the Lovelock-Lifshitz black branes in terms of the temperature
and entropy and find that it is the same in Einstein \cite{Peet2} and Lovelock gravities. We also
consider the thermodynamics of the solutions. We finish our paper with some concluding remarks.

\section{Field Equations}\label{Lag}

The bulk action of third order Lovelock gravity \cite{Lov} in the presence of an
Abelian massive vector field $A^{\mu }$ may be written as
\begin{eqnarray}
I_{\mathrm{bulk}} &=&\frac{1}{16\pi }\int_{\mathcal{M}}d^{n+1}x\sqrt{-g}%
\left( \mathcal{L}_{g}+\mathcal{L}_{m}\right) ,  \nonumber \\
\mathcal{L}_{g} &=&\mathcal{L}_{1}+\alpha _{2}\mathcal{L}_{2}+\alpha _{3}%
\mathcal{L}_{3}-2\Lambda ,  \nonumber \\
\mathcal{L}_{m} &=&-\frac{1}{4}F_{\mu \nu }F^{\mu \nu }-\frac{1}{2}%
m^{2}A_{\mu }A^{\mu }  \label{Act1}
\end{eqnarray}
where $F_{\mu \nu }=\partial _{\lbrack \mu }A_{\nu \rbrack}$, $\Lambda $ is the
cosmological constant, $\alpha _{2}$ and $\alpha _{3}$ are second and third
order Lovelock coefficients, $\mathcal{L}_{1}=R$ is the Einstein-Hilbert
Lagrangian, $\mathcal{L}_{2}=R_{\mu \nu \gamma \delta }R^{\mu \nu \gamma
\delta }-4R_{\mu \nu }R^{\mu \nu }+R^{2}$ is the second order Lovelock
(Gauss-Bonnet) Lagrangian, and

\begin{eqnarray}
\mathcal{L}_{3} &=&R^{3}+2R^{\mu \nu \sigma \kappa }R_{\sigma \kappa \rho
\tau }R_{\phantom{\rho \tau }{\mu \nu }}^{\rho \tau }+8R_{\phantom{\mu
\nu}{\sigma \rho}}^{\mu \nu }R_{\phantom {\sigma \kappa} {\nu \tau}}^{\sigma
\kappa }R_{\phantom{\rho \tau}{ \mu \kappa}}^{\rho \tau }+24R^{\mu \nu
\sigma \kappa }R_{\sigma \kappa \nu \rho }R_{\phantom{\rho}{\mu}}^{\rho }
\nonumber \\
&&+3RR^{\mu \nu \sigma \kappa }R_{\sigma \kappa \mu \nu }+24R^{\mu \nu
\sigma \kappa }R_{\sigma \mu }R_{\kappa \nu }+16R^{\mu \nu }R_{\nu \sigma
}R_{\phantom{\sigma}{\mu}}^{\sigma }-12RR^{\mu \nu }R_{\mu \nu }  \label{L3}
\end{eqnarray}
is the third order Lovelock Lagrangian. We assume that the Gauss-Bonnet
coefficient, which has the dimension of (length)$^{2}$, is positive as in
heterotic string theory \cite{Boul}. In Lovelock gravity only terms with
order less than $[(n+1)/2]$ (where $[x]$ is the integer part of $x$)
contribute to the field equations, the rest being total derivatives in the
action. For third order Lovelock gravity we therefore consider $(n+1)$%
-dimensional spacetimes with $n\geq 6$ (though in situations where we set $%
\hat{\alpha}_{3}=0$ our solutions will be valid for $n\geq 4$).

We write the spherically symmetric gauge field and metric  of an $(n+1)$-dimensional asymptotically Lifshitz static spacetime with zero curvature boundary as
\begin{eqnarray}
A &=&qe^{H(r)}dt=q\frac{r^{z}}{l^{z}}h(r)dt,  \label{A1}\\
ds^{2}&=&-e^{2F(r)}dt^{2}+e^{2G(r)}dr^{2}+l^2 e^{2R(r)}\sum%
\limits_{i=1}^{n-1}(dx^{i})^{2}  \label{met1} \\
&=&-\frac{r^{2z}}{l^{2z}}f(r)dt^{2}+\frac{l^{2}dr^{2}}{r^{2}g(r)}%
+r^{2}\sum\limits_{i=1}^{n-1}(dx^{i})^{2},  \label{met2}
\end{eqnarray}
where the new metric functions are related to our previous notation \cite{DM}
through the following relations
\begin{eqnarray}
F(r) &=&\frac{1}{2}\ln f(r)+z\ln \frac{r}{l},  \nonumber \\
G(r) &=&-\frac{1}{2}\ln g(r)-\ln \frac{r}{l},  \nonumber \\
R(r) &=&\ln \frac{r}{l},  \nonumber \\
H(r) &=&\ln h(r)+z\ln \frac{r}{l}  \label{func}
\end{eqnarray}
and  in Eq. (\ref{met2}) we have chosen coordinates by taking $R(r)=\ln (r/l)$.

Our previous study of Lovelock-Lifshitz black holes \cite{DM} showed that if
\begin{eqnarray}
m^{2} &=&\frac{(n-1)z}{l^{2}},\text{ \ \ \ \ }q^{2}=\frac{2(z-1)L^{4}}{zl^{4}%
},  \nonumber \\
\Lambda  &=&-\frac{[(z-1)^{2}+n(z-2)+n^{2}]L^{4}+n(n-1)(\hat{\alpha}%
_{2}l^{2}-2\hat{\alpha}_{3})}{2l^{6}},  \label{Coef}
\end{eqnarray}
where $L^{4}=l^{4}-2\hat{\alpha}_{2}l^{2}+3\hat{\alpha}_{3}$, then the
action (\ref{Act1}) supports solutions asymptotic to the Lifshitz solution \cite{DM}
\begin{equation}
ds^{2}=-\frac{r^{2z}}{l^{2z}}dt^{2}+\frac{l^{2}dr^{2}}{r^{2}}%
+r^{2}\sum\limits_{i=1}^{n-1}dx_{i}^{2}.  \label{metLif}
\end{equation}

Since we are assuming spherical symmetry, we can reduce the action to one
dimension and subsequently obtain the equations of motion.
After integration by parts, the one dimensional Lagrangian may be
written as $\mathcal{L}_{1D}=l^{n-1}(\mathcal{L}_{1g}+\mathcal{L}_{1m})$, where
\begin{eqnarray}
\mathcal{L}_{1g} &=&(n-1)\Big\{-2\frac{\Lambda }{n-1}e^{2G}+\left[ 2F^{\prime
}R^{\prime }+(n-2)R^{\prime 2}\right] -\frac{\hat{\alpha}_{2}}{3}\left[
4F^{\prime }R^{\prime 3}+(n-4)R^{\prime 4}\right] e^{-2G} \nonumber\\
&&+\frac{\hat{\alpha}_{3}}{5}\left[ 6F^{\prime }R^{\prime 5}+(n-6)R^{\prime
6}\right] e^{-4G}\Big\}e^{F-G+(n-1)R} \nonumber\\
\mathcal{L}_{1m} &=&\frac{1}{2}q^{2}\left( m^{2}+H^{\prime 2}e^{-2G}\right)
e^{-F+G+(n-1)R+2H}, \label{Act2}
\end{eqnarray}
where prime denotes the derivative with respect to $r$, and we define $\hat{%
\alpha}_{2}\equiv (n-2)(n-3)\alpha _{2}$ and $\hat{\alpha}_{3}\equiv
(n-2)...(n-5)\alpha _{3}$ for convenience. The equations of motion following
from this action may be written as:
\begin{eqnarray}
\mathcal{L}_{1g}-\mathcal{L}_{1m}&=&\left\{ 2(n-1)\left( R^{\prime }-\frac{2%
}{3}\hat{\alpha}_{2}R^{\prime 3}e^{-2G}+\frac{3}{5}\hat{\alpha}_{3}R^{\prime
5}e^{-4G}\right) e^{F-G+(n-1)R}\right\} ^{\prime },  \label{e1} \\
\mathcal{L}_{1g}+\mathcal{L}_{1m} &=& \Big\{ 2\Big[ F^{\prime
}+(n-2)R^{\prime }-\frac{2}{3}\hat{\alpha}_{2}\left( 3F^{\prime }R^{\prime
2}+(n-4)R^{\prime 3}\right) e^{-2G}\\
&& +\frac{3}{5}\hat{\alpha}_{3}\left(
5F^{\prime }R^{\prime 4}+(n-6)R^{\prime 5}\right) e^{-4G}\Big]
e^{F-G+(n-1)R}\Big\} ^{\prime },  \label{e2} \\
2\mathcal{L}_{1m} &=&\left\{ q^{2}H^{\prime }e^{-F-G+(n-1)R+2H}\right\}
^{\prime },  \label{e3} \\
0 &=&\left[ 2F^{\prime }R^{\prime }+(n-2)R^{\prime 2}\right]-\hat{%
\alpha}_{2}\left[ 4F^{\prime }R^{\prime 3}+(n-4)R^{\prime 4}\right] e^{-2G}
\nonumber \\
&&+\hat{\alpha}_{3}\left[ 6F^{\prime }R^{\prime 5}+(n-6)R^{\prime 6}\right]
e^{-4G}+\frac{2\Lambda }{n-1}e^{2G} \nonumber\\
&& -\frac{q^{2}}{2(n-1)}\left(
m^{2}e^{2G}-H^{\prime 2}\right) e^{2(H-F)}.  \label{e4}
\end{eqnarray}
Now subtracting \ the summation of Eqs. (\ref{e1}) and (\ref{e3}) from Eq. (%
\ref{e2}) one obtains:
\[
\left\{ 2(F^{\prime }-R^{\prime })\left( 1-2\hat{\alpha}_{2}R^{\prime
2}e^{-2G}+3\hat{\alpha}_{3}R^{\prime 4}e^{-4G}\right)
e^{F-G+(n-1)R}-q^{2}H^{\prime }e^{-F-G+(n-1)R+2H}\right\} ^{\prime }=0,
\]
which shows that
\begin{eqnarray}
\mathcal{C}_{0} &=&2(F^{\prime }-R^{\prime })\left( 1-2\hat{\alpha}%
_{2}R^{\prime 2}e^{-2G}+3\hat{\alpha}_{3}R^{\prime 4}e^{-4G}\right)
e^{F-G+(n-1)R}-q^2 H^{\prime }e^{-F-G+(n-1)R+2H}  \nonumber \\
&=&\left\{ \left( 1-2\frac{\hat{\alpha}_{2}}{l^{2}}g+3\frac{\hat{\alpha}_{3}%
}{l^{4}}g^{2}\right) \left[ rf^{\prime }+2(z-1)f\right] -q^{2}(zh+rh^{\prime
})h\right\} \frac{r^{n+z-1}}{l^{z+1}}\left( \frac{f}{g}\right) ^{1/2},
\label{Cons}
\end{eqnarray}
is conserved along the radial coordinate $r$. One may note that the action (%
\ref{Act2}) and constant (\ref{Cons}) reduce to those of Ref. \cite{Peet1}
for $n=3$ and $\hat{\alpha}_{2}=\hat{\alpha}_{3}=0$. This conserved quantity
is associated with the shift
\begin{equation}
\begin{pmatrix}
F(r) \\
R(r) \\
G(r) \\
H(r)
\end{pmatrix}
\rightarrow
\begin{pmatrix}
F(r)+\delta  \\
R(r)-\frac{\delta }{n-1} \\
G(r) \\
H(r)+\delta
\end{pmatrix}
,  \label{shift}
\end{equation}
where $\delta $ is a constant, and reduces to the diffeomorphism introduced
in \cite{Peet1} for $n=3$ and $\hat{\alpha}_{2}=\hat{\alpha}_{3}=0$.

We pause to remark that for $z=1$ with $f(r)=g(r)$, the constant (\ref{Cons}) reduces to
\begin{eqnarray*}
\mathcal{C}_{0} &=&\frac{r^{n+1}}{l^{2}}\left( 1-2\frac{\hat{\alpha}_{2}}{%
l^{2}}f+3\frac{\hat{\alpha}_{3}}{l^{4}}f^{2}\right) f^{\prime } \\
&=&\frac{r^{n+1}}{l^{2}}\left(f-\frac{\hat{\alpha}_{2}}{l^{2}}f^{2}+\frac{%
\hat{\alpha}_{3}}{l^{4}}f^{3}\right) ^{\prime },
\end{eqnarray*}
which is known to be constant in third order Lovelock gravity and  is
proportional to the mass parameter of the spacetime \cite{DP}.

\section{Finite Action for Lovelock-Lifshitz Solutions}\label{finite}

In Ref. \cite{Ross}, the authors define a finite action for Lifshitz theory
in 4-dimensional Einstein gravity which satisfies $\delta I=0$ with
appropriate boundary conditions by adding appropriate local counterterms to
the action (\ref{Act1}). In this section we generalize this action to the case
of ($n+1)$-dimensional Lovelock gravity. As in the case of Einstein gravity,
to preserve the diffeomorphism invariance of the action  the counterterms rendering the action finite
and yielding a well-defined variational
principle should be covariant in the boundary fields. We consider $I=I_{%
\mathrm{bulk}}+I_{\mathrm{bdy}}$, where $I_{\mathrm{bulk}}$ is given in Eq. (%
\ref{Act1}) and $I_{\mathrm{bdy}}$ is the sum of the boundary terms which
are needed to have a well-defined variational principle and the counterterms
which guarantees the finiteness of the action. $I_{\mathrm{bdy}}$, for the
case of zero curvature boundary which is our interest, may be written as
\begin{eqnarray}
I_{\mathrm{bdy}}&=&\frac{1}{8\pi }\int_{\partial \mathcal{M}}d^{n}x\sqrt{-h}%
\Big\{ K-\frac{n-1}{l}+2\alpha _{2}\left( J-\frac{a(n)}{l^{3}}\right) \nonumber\\
&& +3\alpha _{3}\left( P-\frac{b(n)}{l^{5}}\right)+\frac{1}{2}f(A_{\alpha
}A^{\alpha })\Big\} +I_{\mathrm{deriv}},  \label{Ibdy}
\end{eqnarray}
where the boundary $\partial \mathcal{M}$ is the hypersurface at some
constant $r$, $h_{\alpha \beta }$ is the induced metric, $K$ is the trace of
the extrinsic curvature, $K_{\alpha \beta }=\nabla _{(\alpha }n_{\beta )}$
of the boundary (where the unit vector $n^{\mu }$ is orthogonal to the
boundary and outward-directed), $a(n)$ and $b(n)$ are two dimensionless
constants  depending on $n$, and $J$ and $P$ are the traces of \cite{DM2}
\begin{eqnarray}
J_{\alpha \beta} &=&\frac{1}{3}%
(2KK_{\alpha \gamma}K_{ \beta}^{ \gamma}+K_{\gamma  \delta} K^{\gamma  \delta} K_{\alpha \beta}-2K_{\alpha \gamma}K^{\gamma  \delta} K_{\delta  \beta}-K^{2}K_{\alpha \beta})
\label{Jab} \\
P_{\alpha \beta} &=&\frac{1}{5}%
\Big\{[K^{4}-6K^{2}K^{\gamma  \delta} K_{\gamma  \delta} +8KK_{\gamma  \delta} K_{ \epsilon} ^{\delta }K^{\epsilon   \gamma}-6K_{\gamma  \delta} K^{\delta \epsilon} K_{\epsilon \psi}K^{\psi \gamma}+3(K_{\gamma  \delta} K^{\gamma  \delta} )^{2}]K_{\alpha \beta}
\nonumber \\
&&-(4K^{3}-12KK_{\epsilon \delta} K^{\epsilon   \delta} +8K_{\delta  \epsilon} K_{\psi}^{ \epsilon} K^{\psi \delta} )K_{\alpha \gamma}K_{ \beta}^{ \gamma}-24KK_{\alpha \gamma}K^{\gamma \delta} K_{\delta  \epsilon} K_{ \beta}^{ \epsilon}
\nonumber \\
&&+(12K^{2}-12K_{\epsilon   \psi}K^{\epsilon   \psi})K_{\alpha \gamma}K^{\gamma  \delta} K_{\delta  \beta}+24K_{\alpha \gamma}K^{\gamma  \delta} K_{\delta  \epsilon} K^{\epsilon \psi}K_{\beta \psi}\Big\}
\label{Pab}
\end{eqnarray}
In Eq. (\ref{Ibdy}), $I_{\mathrm{deriv}}$ is a collection of terms involving
derivatives of the boundary fields, which could involve both the curvature
tensor constructed from the boundary metric and covariant derivatives of $%
A_{\alpha }$, Since the boundary is flat and the fields are constants for
\ref{metLif}, this term will not contribute to the on-shell
value of the action for the pure Lifshitz solution or its first variation
around the Lifshitz background and therefore we ignore it throughout the
paper. The matter part of the action (\ref{Ibdy}) is the same as the matter
part of the action in Einstein gravity. Thus, as in Ref. \cite{Ross}, an
arbitrary function $f(A^{\alpha }A_{\alpha })$ is added to the action which is
due to the fact that on the boundary $A_{\alpha }A^{\alpha }=-q^{2}$ is
constant for Lifshitz solutions.

The variation of the action about a solution of the equations of motion is
just the boundary term,
\begin{eqnarray}
\delta I &=&\frac{1}{16\pi }\int_{\partial \mathcal{M}}d^{n}x\sqrt{-h}\Big\{
\Pi _{\alpha \beta }\delta h^{\alpha \beta }-n^{\mu }F_{\mu \nu }\delta
A^{\nu }  \\
&& +f^{\prime }(A_{\alpha }A^{\alpha })(2A_{\alpha }\delta A^{\alpha
}+A_{\alpha }A_{\beta }\delta h^{\alpha \beta })-\frac{1}{2}f(A_{\alpha
}A^{\alpha })h_{\alpha \beta }\delta h^{\alpha \beta }\Big\} ,  \nonumber
\end{eqnarray}
where $\Pi _{\alpha \beta }=\Pi _{\alpha \beta }^{(1)}+\Pi _{\alpha \beta
}^{(2)}+\Pi _{\alpha \beta }^{(3)}$, with
\begin{eqnarray*}
\Pi _{\alpha \beta }^{(1)} &=&K_{\alpha \beta }-Kh_{\alpha \beta }+\frac{n-1%
}{l}h_{\alpha \beta }, \\
\Pi _{\alpha \beta }^{(2)} &=&2\alpha _{2}\left( 3J_{\alpha \beta
}-Jh_{\alpha \beta }+\frac{a(n)}{l^{3}}h_{\alpha \beta }\right) , \\
\Pi _{\alpha \beta }^{(3)} &=&3\alpha _{3}\left( 5P_{\alpha \beta
}-Ph_{\alpha \beta }+\frac{b(n)}{l^{5}}h_{\alpha \beta }\right) .
\end{eqnarray*}
As in the case of Einstein gravity \cite{Ross}, one has for the
Lovelock-Lifshitz spacetime (\ref{metLif}) $n^{\mu }F_{\mu \nu }\delta A^{\nu
}=zqr^{z}/l^{z}\delta A^{t}$,
\[
\Pi _{tt}^{(1)}=0,\text{ \ \ }\Pi _{ij}^{(1)}=-\frac{z-1}{l}r^{2}\delta
_{ij},
\]
and it is a matter of calculation to show that if one chooses
\begin{eqnarray*}
a(n) &=&-(n-1)(n-2)(n-3), \\
b(n) &=&(n-1)...(n-5),
\end{eqnarray*}
then one also has
\begin{eqnarray*}
\Pi _{tt}^{(2)} &=&0,\text{ \ \ }\Pi _{ij}^{(2)}=\frac{2(z-1)\hat{\alpha}_{2}%
}{l^{3}}r^{2}\delta _{ij}, \\
\Pi _{tt}^{(3)} &=&0,\text{ \ \ }\Pi _{ij}^{(3)}=-\frac{3(z-1)\hat{\alpha}%
_{3}}{l^{5}}r^{2}\delta _{ij}.
\end{eqnarray*}
That is,
\begin{equation}
\Pi _{tt}=0,\text{ \ \ \ \ }\Pi _{ij}=\frac{(1-z)L^{4}}{l^{5}}r^{2}\delta
_{ij}. \label{Pis}
\end{equation}

Equation (\ref{Pis}) shows that, as in the case of Einstein gravity \cite{Ross}, there are variations involving $\delta h_{ij}$
and $\delta A^t$ that we need to cancel. Using the same argument as in \cite{Ross}, the finite action may be written as
\begin{eqnarray}
I &=&\frac{1}{16\pi }\int_{\mathcal{M}}d^{n+1}x\sqrt{-g}(\mathcal{L}_{g}+%
\mathcal{L}_{m})  \nonumber \\
&&+\frac{1}{8\pi }\int_{\partial \mathcal{M}}d^{n}x\sqrt{-h}(K+2\alpha
_{2}J+3\alpha _{3}P-\frac{(n-1)L^{4}}{l^{5}}-\frac{zq}{2l}\sqrt{-A_{\alpha
}A^{\alpha }})  \label{FinAct}
\end{eqnarray}
It is remarkable that fixing a single coefficient suffices to cancel both
divergences associated with $\delta h^{ij}$ and $\delta A^{t}$. Note
that when $z=1$, then $q=0$ and these definitions reduce to the familiar AdS
rules for third order Lovelock gravity \cite{DM2}. If one defines
\begin{equation}
S_{\alpha \beta }=\frac{\sqrt{-h}}{16\pi }\left[ \Pi _{\alpha \beta }+\frac{%
zq}{2l}(-A_{\alpha }A^{\alpha })^{-1/2}(A_{\alpha }A_{\beta }-A_{\gamma
}A^{\gamma }h_{\alpha \beta })\right] ,  \label{Sab}
\end{equation}
\begin{equation}
S_{\alpha }=\frac{\sqrt{-h}}{16\pi }[n^{\mu }F_{\mu \alpha }+z\alpha
(-A_{\alpha }A^{\alpha })^{-1/2}A_{\alpha }],  \label{Sa}
\end{equation}
then the general variation of the action is
\begin{equation}
\delta I=\int d^{n}x(S_{\alpha \beta }\delta h^{\alpha \beta }+S_{\alpha
}\delta A^{\alpha }).
\end{equation}
In the background (\ref{metLif}), we have $S_{\alpha \beta }=0$, $S_{\alpha }=0
$ due to cancellations between the different terms, and this action
satisfies $\delta I=0$ for arbitrary variations around (\ref{metLif}). Thus,
we have a finite on-shell action which defines a well-defined variational
principle for our background spacetime.

In the holographic renormalization programme for  gauge-gravity duality in
relativistic field theories, one computes the finite stress  tensor after constructing a well-defined finite action, since it carries important physical information about the dual field theory. This job has been done for asymptotically AdS spacetimes \cite{Henn,Bal}. For
asymptotically Lifshitz spacetimes, the dual field theory is
nonrelativistic, so it will not have a covariant relativistic stress
tensor. However  one can define a  stress tensor complex \cite{Ross}, consisting of the energy density $\mathcal{E}{}$, energy flux ${}\mathcal{E}_{i}$, momentum density ${}\mathcal{P}_{i}$ and
spatial stress tensor $\mathcal{P}_{ij}$, satisfying the conservation
equations
\begin{equation}
\partial _{t}{}\mathcal{E}{}+\partial _{i}{}\mathcal{E}{}^{i}=0,\quad
\partial _{t}{}\mathcal{P}_{j}+\partial _{i}\mathcal{P}_{j}^{i}=0, \label{Conserv}
\end{equation}
where
\begin{equation}
\mathcal{E}{}{}=2S_{\ t}^{t}-S^{t}A_{t},\quad {}\mathcal{E}{}^{i}=2S_{\
t}^{i}-S^{i}A_{t},  \label{En}
\end{equation}
and
\begin{equation}
{}\mathcal{P}_{i}=-2S_{\ i}^{t}+S^{t}A_{i}\quad \mathcal{P}_{i}^{j}=-2S_{\
i}^{j}+S^{j}A_{i}. \label{Pi}
\end{equation}
where in the Lovelock case  $S_{\alpha \beta}$
and $S_{\alpha}$ are given by (\ref{Sab},\ref{Sa}) respectively. Consequently the stress tensor complex (\ref{En})and (\ref{Pi}) satisfies the conservation
Eqs. (\ref{Conserv}).

\section{The Constant $\mathcal{C}_{0}$}\label{Constant}

In this section, we want to calculate the constant $\mathcal{C}_{0}$, which
conserved along the radial coordinate $r$. Since there is no exact Lovelock-Lifshitz solution
(except under special circumstances),
we calculate it at the horizon and at infinity. We will use this to relate
the constant that appears in the expansion at $r=\infty $ to the coefficients at the horizon.
\subsection{$\mathcal{C}_{0}$ at the horizon}

We begin by first reviewing the results found in \cite{DM} for the expansion
near the horizon. Requiring that $f(r)$ and $g(r)$ go to zero linearly, that
is
\begin{eqnarray}
f(r) &=&f_{1}\left\{
(r-r_{0})+f_{2}(r-r_{0})^{2}+f_{3}(r-r_{0})^{3}+f_{4}(r-r_{0})^{4}+...\right%
\} ,  \nonumber \\
g(r)
&=&g_{1}(r-r_{0})+g_{2}(r-r_{0})^{2}+g_{3}(r-r_{0})^{3}+g_{4}(r-r_{0})^{4}+...,
\nonumber \\
h(r) &=&f_{1}^{1/2}\left\{
h_{0}+h_{1}(r-r_{0})+h_{2}(r-r_{0})^{2}+h_{3}(r-r_{0})^{3}+h_{4}(r-r_{0})^{4}+...\right\} ,
\label{ExpH}
\end{eqnarray}
and inserting these expansions into the equations of motion arising from the
action (\ref{Act1}) for the metric (\ref{met2}) with the conditions (\ref{Coef})
[Eqs (2.8)-(2.10) of Ref.
\cite{DM}], and solving for the various
coefficients, we find that $h_{0}=0$. This is consistent with the fact that
the flux $dA$ should go to a constant at the horizon.

By rescaling the time coordinate, we can adjust the constant $f_{1}$\ by an overall
multiplicative factor (note the use of $f_{1}^{1/2}$\ in the expansion of $%
h(r)$\ as well, which is due to $dt$\ in the one-form $A$). It is straightforward to
find all the coefficients in terms of the two constants $r_{0}$ and $h_{1}$.

Using the above expansion, we evaluate the constant $\mathcal{C}_{0}$ (%
\ref{Cons}) at $r=r_{0}$:
\begin{equation}
\mathcal{C}_{0}=\frac{r_{0}^{n+z}\sqrt{f_{1}g_{1}}}{l^{z+1}}.  \label{D0r0}
\end{equation}
This must be preserved along the flow in $r$.

\subsection{$\mathcal{C}_{0}$ at infinity}

We now turn to the calculation of $\mathcal{C}_{0}$\ at large $r$. In order
to do this, we investigate the behavior of the metric functions at large $r$
by using straightforward perturbation theory:
\begin{eqnarray*}
f(r) &=&1+\varepsilon f_{1}(r), \\
g(r) &=&1+\varepsilon g_{1}(r), \\
h(r) &=&1+\varepsilon h_{1}(r),
\end{eqnarray*}
and finding the field equations up to the first order in $\varepsilon $. We
obtain
\begin{eqnarray}
0 &=&2r^{2}h_{1}^{\prime \prime }+2(n+z)rh_{1}^{\prime }+zr\left(
g_{1}^{\prime }-f_{1}^{\prime }\right) +2(n-1)zg_{1},  \nonumber \\
0 &=&2(z-1)rh_{1}^{\prime }+(n-1)rg_{1}^{\prime }+\left[ z(z-1)+n(n-1)\right]
g_{1}-(z-1)(n+z-1)(f_{1}-2h_{1}),  \nonumber \\
0 &=&2(z-1)rh_{1}^{\prime }+(n-1)rg_{1}^{\prime }+\left[
z(z-1)+n(n-1)+2(n-1)(z-1)\mathcal{B}\right] g_{1}  \nonumber \\
&&-(z-1)(z-n+1)(f_{1}-2h_{1})  \label{infinity}
\end{eqnarray}
where $\mathcal{B}=(l^{4}-4\hat{\alpha}_{2}l^{2}+9\hat{\alpha}_{3})/L^{4}$.
Note that all the parameters of Lovelock gravity are in $\mathcal{B}$, with $%
\mathcal{B}=1$ in Einstein gravity.

The solution of Eqs. (\ref{infinity}) is
\begin{eqnarray}
h_{1}(r) &=&-\frac{C_{1}}{r^{n+z-1}}-\frac{C_{2}}{r^{(n+z-1+\gamma )/2}}-%
\frac{C_{3}}{r^{(n+z-1-\gamma )/2}},  \nonumber \\
f_{1}(r) &=&-\frac{C_{1}F_{1}}{r^{n+z-1}}-\frac{C_{2}F_{2}}{r^{(n+z-1+\gamma
)/2}}-\frac{C_{3}F_{3}}{r^{(n+z-1-\gamma )/2}},  \nonumber \\
g_{1}(r) &=&-\frac{C_{1}G_{1}}{r^{n+z-1}}-\frac{C_{2}G_{2}}{r^{(n+z-1+\gamma
)/2}}-\frac{C_{3}G_{3}}{r^{(n+z-1-\gamma )/2}},  \label{Finf}
\end{eqnarray}
where
\begin{eqnarray*}
\gamma &=&\left\{ (17-8\mathcal{B})z^{2}-2(3n+9-8\mathcal{B})z+n^{2}+6n+1-8\mathcal{B}%
\right\} ^{1/2}, \\
F_{1} &=&2\left( z-1\right) \left( z-n+1\right) {\mathcal{K}}^{-1}, \\
F_{2} &=&\left( \mathcal{F}_{1}-\mathcal{F}_{2}\right) \left\{ 8z\mathcal{K}%
\left[ (z-1)\mathcal{B}+2n+z-3\right] \right\} ^{-1}, \\
F_{3} &=&\left( \mathcal{F}_{1}+\mathcal{F}_{2}\right) \left\{ 8z\mathcal{K}%
\left[ (z-1)\mathcal{B}+2n+z-3\right] \right\} ^{-1}, \\
G_{1} &=&2\left( z-1\right) \left( n+z-1\right) {\mathcal{K}}^{-1}, \\
G_{2} &=&\left( \mathcal{G}_{1}+\mathcal{G}_{2}\right) \left\{ 8z\mathcal{K}%
\left[ (z-1)\mathcal{B}+2n+z-3\right] \right\} ^{-1}, \\
G_{3} &=&\left( \mathcal{G}_{1}-\mathcal{G}_{2}\right) \left\{ 8z\mathcal{K}%
\left[ (z-1)\mathcal{B}+2n+z-3\right] \right\} ^{-1}, \\
\mathcal{K} &=&(z-1)(n+z-1)\mathcal{B}+z(z-1)+n(n-1), \\
\mathcal{F}_{1} &=&8(z-1)[(z-1)(n+z-1)\mathcal{B}+z(z-1)+n(n-1)] \\
&&\times \lbrack (z-1)(n+3z-3)\mathcal{B}-2z^{2}+(n+3)z+n(n-2)-1], \\
\mathcal{F}_{2} &=&\gamma \lbrack n-1+(z-1)\mathcal{B}]\Big\{8(1+\mathcal{B}%
)(z-1)^{3} \\
&&+(17n-9+8\mathcal{B})(z-1)^{2}+2(n+8)(n-1)(z-1)+n^{2}(n-1)-(n-1)\gamma ^{2}%
\Big\}, \\
\mathcal{G}_{1} &=&8(z-1)[2(z-1)\mathcal{B}-3z+3n-1][(z-1)(n+z-1)\mathcal{B}%
+z(z-1)+n(n-1)], \\
\mathcal{G}_{2} &=&\Big\{8(1+\mathcal{B})(z-1)^{3}+(17n-9+8\mathcal{B}%
)(z-1)^{2} \\
&&+2(n+8)(n-1)(z-1)+n^{2}(n-1)\Big\}\gamma -(n-1)\gamma ^{3}.
\end{eqnarray*}

The case of $z=n-1$ in Einstein gravity needs special consideration. In this
case $\gamma =2z$, and therefore the second term in Eqs. (\ref{Finf}) is
exactly the same as the first term. Also, one should choose $C_{3}=0$ in
order to have   suitable asymptotically behavior for the functions. It is a matter of
calculation to show that the solution of Eqs. (\ref{infinity}) for $z=n-1$, together with
the constraint that these functions should go to zero for large $r$,
may be written as
\begin{eqnarray}
h_{1}(r) &=&-\frac{C_{1}+C_{2}\ln r}{r^{2(n-1)}},  \nonumber \\
f_{1}(r) &=&-\frac{3n-4}{(n-1)(2n-3)}\frac{C_{2}}{r^{2(n-1)}},  \nonumber \\
g_{1}(r) &=&-\frac{2(n-2)(C_{1}+C_{2}\ln r)}{(2n-3)r^{2(n-1)}}-\frac{%
(n^{2}-2)C_{2}}{(n-1)(2n-3)^{2}r^{2(n-1)}},  \label{FinfSpz}
\end{eqnarray}
where $C_{1}$ and $C_{2}$ are integration constants.

Now we want to calculate the conserved quantity $\mathcal{C}_{0}$.  We restrict ourselves to the case
that $\gamma \geq n+z-1$, which for Einstein gravity means that $z \geq n-1$. However in Lovelock gravity, this condition holds provided $z\geq(n-\mathcal{B})/(2-\mathcal{B})$,
where $\mathcal{B}$ depends on the Lovelock coefficients. Note that
this condition does not hold for $\mathcal{B} \geq 2$.
If $z\geq(n-\mathcal{B})/(2-\mathcal{B})$ then $C_{3}=0$, since at large $r$ the functions $f_{1}(r)$, $g_{1}(r)$ and $h_{1}(r)$ should go to zero as $r$ goes to infinity, and the contribution in
$\mathcal{C}_{0}$ from $C_{2}$ is zero. Thus, one has only the first terms
in the expansions (\ref{Finf}) and the constant $\mathcal{C}_{0}$ can be
obtained as
\begin{equation}
\mathcal{C}_{0}=\frac{2(z-1)(z+n-1)^{2}\{(z-n+1)l^{4}+2(n-2)\hat{\alpha}%
_{2}l^{2}-3(n+z-3)\hat{\alpha}_{3}\}}{zl^{z+5}\mathcal{K}}C_{1}.
\label{D0inf}
\end{equation}
We also find that the terms due to the second order perturbation do
not  contribute to the conserved quantity (\ref{D0inf}).

The above constant (\ref{D0inf}) in Einstein gravity for $z=n-1$ is zero. In this case we must use the
expansion (\ref{FinfSpz}) to obtain the conserved quantity $\mathcal{C}_{0}$, which gives
\begin{equation}
\mathcal{C}_{0}=\frac{4(n-1)}{(2n-3)l^{n}}C_{2},\text{ \ \ (}z=n-1\text{,
Einstein)}  \label{D0infSpz}
\end{equation}

\section{Black Brane thermodynamics}\label{Therm}

The entropy of a black hole in Lovelock gravity is \cite{Wald}
\begin{equation}
S=\frac{1}{4}\sum_{k=1}^{p}k\alpha _{k}\int d^{n-1}x\sqrt{\tilde{g}}\tilde{%
\mathcal{L}}_{k-1},  \label{Enta}
\end{equation}
where the integration is done on the $(n-1)$-dimensional spacelike
hypersurface of the Killing horizon with induced metric $\tilde{g}_{\mu \nu
} $ (whose determinant is $\tilde{g}$), and $\tilde{\mathcal{L}}_{k}$ is the
$k $th order Lovelock Lagrangian of $\tilde{g}_{\mu \nu }$. The entropy of a
black brane per unit volume of the horizon in third order Lovelock gravity
is
\begin{equation}
S=\frac{1}{4}r_{0}^{n-1}.  \label{Ent}
\end{equation}

The temperature of the event horizon can be obtained by using the expansion
(\ref{ExpH}), yielding
\begin{equation}
T=\frac{r_{0}^{z+1}\sqrt{f_{1}g_{1}}}{4\pi l^{z+1}}.  \label{Temp}
\end{equation}
Using the value of $\mathcal{C}_{0}$\ at horizon (\ref{D0r0}) and Eqs. (%
\ref{Ent}) and (\ref{Temp}), we find
\begin{equation}
\mathcal{C}_{0}=16\pi TS.  \label{D0H}
\end{equation}
Alternatively, one can calculate the energy density of the black brane by
using Eqs. (\ref{Sab}), (\ref{Sa}) and (\ref{En}).  We find that
\begin{equation}
\mathcal{E}=\frac{(n-1)(z-1)(z+n-1)\{(z-n+1)l^{4}+2(n-2)\hat{\alpha}%
_{2}l^{2}-3(n+z-3)\hat{\alpha}_{3}\}}{8\pi zl^{z+5}\mathcal{K}}C_{1}.
\label{Ener}
\end{equation}
For $z=n-1$ the above expression for the energy density in Einstein
gravity vanishes  and we must use the expansion (\ref{FinfSpz}) to find the energy density in Einstein
gravity with $z=n-1$.  This gives
\begin{equation}
\mathcal{E}=\frac{(n-1)}{8\pi (2n-3)l^{n}}C_{2}.  \label{EnerSpz}
\end{equation}

Dividing Eq. (\ref{Ener}) by Eq. (\ref{D0inf}) (or, for $z=n-1$, (\ref{EnerSpz})  by (%
\ref{D0infSpz})) and using Eq. (\ref{D0H}) we find
\begin{equation}
\mathcal{E}=\frac{n-1}{n+z-1}TS.  \label{EnTS}
\end{equation}
which is valid for all $z$.
Note that both the constant $\mathcal{C}_{0}$ and the energy density $%
\mathcal{E}$ are different from their counterparts in Einstein gravity since the coefficient of   $r^{-(n+z-1)}$ at large $r$ for the
functions $f(r)$, $g(r)$ and $h(r)$ differs. However, as one can see
from Eq. (\ref{EnTS}), $\mathcal{E}(T,S)$ is the same in both Einstein \cite{Peet2} and
Lovelock gravity.

For the case of $z=1$, for which the field equations have exact solution, the
energy density, entropy density and temperature are \cite{DM2}
\begin{eqnarray*}
\mathcal{E} &=&\frac{(n-1)m}{16\pi }=\frac{(n-1)r_{0}^{n}}{16\pi l^{2}}, \\
T &=&\frac{nr_{0}}{4\pi l^{2}},\text{ \ \ \ \ \ \ \ \ }S=\frac{1}{4}%
r_{0}^{n-1},
\end{eqnarray*}
and therefore $\mathcal{E}=(n-1)ST/n$,  consistent with Eq. (\ref
{EnTS}) for $z=1$.

Using the first law of thermodynamics $d\mathcal{E}=TdS$ with the
relation (\ref{EnTS}) for the energy density, one obtains
\begin{equation}
\log T=\frac{z}{n-1}\log S+\Gamma ,  \label{logT}
\end{equation}
where $\Gamma $ an the integration constant that depends on the Lovelock
coefficients, $z$ and the dimension of spacetime. It can be found
numerically. This result is consistent with our previous numerical solution
\cite{DM}, which shows that the slope of $\log T$ versus $\log S$ are the
same for Einstein and Lovelock gravity, while $\Gamma $ is different. Furthermore,
by use of Eqs. (\ref{Ent}) and (\ref{logT}), the
temperature is proportional to $r_{0}^{z}$.

\section{Conclusions}

In this paper, we introduced the finite action of third order Lovelock gravity in the
presence of a massive vector field with a flat boundary. Indeed, we generalized the  counterterm method introduced
in \cite{DM2} for asymptotically AdS black branes to the case of asymptotic Lifshitz black branes.
We also defined the finite stress tensor complex, and computed the energy density of the Lovelock-Lifshitz black branes.
We  then used the field equations to find a conserved quantity along the $r$ coordinate. This constant, which is the
generalization of the constant introduced in \cite{Peet1}, has the role of connecting the metric parameters at the horizon and at infinity.
We used these generalizations to investigate the thermodynamics of Lovelock-Lifshitz black branes introduced in \cite{DM}.  We found
that the relationship between the energy density, temperature, and entropy density is unchanged from Einsteinian gravity, even though the subleading large-$r$ behavior of Lovelock-Lifshitz black branes is different from the Einsteinian Lifshitz solutions.  We made use of the first law of thermodynamics to obtain the relationship between entropy and temperature.  We found this is also the same as the Einsteinian case, apart from a constant of integration that depends on the parameters in the Lovelock action. These results are consistent with the numerical analysis of Lovelock-Lifshitz black branes investigated in \cite{DM}.

Our counterterm method can be applied only to the case of $k=0$ solutions. It would be interesting to generalize this method to the $k=\pm 1$ cases. Its generalization  to  quasitopological gravity with cubic-curvature terms that are not supersymmetric and therefore is different from the third order Lovelock gravity in the context of holography \cite{Myers} would also be of interest. It is known that the third order Lovelock term does not contribute to the three-point functions of gravitons in flat space \cite{Met}, while the corresponding contribution in the AdS background does not vanish \cite{Boer}. Employing our action to calculate
two and three-point functions in the pure Lifshitz background, and considering their differences with respect to
the case of asymptotic flat and AdS solutions is another topic worth investigating.

\section*{Acknowledgements}
This work was supported by the Natural Sciences and Engineering Research Council of Canada and
the Research Council of Shiraz University.

\end{document}